\documentclass[12pt]{iopart}
\usepackage{graphicx}

\usepackage{iopams}
\expandafter\let\csname equation*\endcsname\relax
\expandafter\let\csname endequation*\endcsname\relax
\usepackage{amsmath}

\usepackage{amsfonts,amsthm,bm,mathtools}%,environ}

\usepackage[usenames,dvipsnames]{color}
\usepackage{url}
\usepackage{epstopdf}
\usepackage{mathtools} % text over arrows

\usepackage{cite}

\usepackage{pdfpages}

\definecolor{linkcolor}{rgb}{0,0,0.6} %hyperlink
\usepackage[colorlinks=true, pdfstartview=FitV, linkcolor= linkcolor, citecolor= linkcolor, urlcolor= linkcolor, hyperindex=true,hyperfigures=true]{hyperref}

\newcommand{\tder}{\dot}
\newcommand{\non}{\nonumber \\}
\newcommand{\toprule}{\br}
\newcommand{\bottomrule}{\br}
\newcommand{\midrule}{\mr}
\newcommand{\rev}[1]{{\bf{ \textcolor{Blue}{#1}}}}%
\renewcommand{\rev}[1]{#1} %  CLEANED FROM COLOR AND BF

\begin{document}

\title[]{Modelling the deceleration of COVID-19 spreading}

\author{Giacomo Barzon$^{*}$}
\author{Karan Kabbur Hanumanthappa Manjunatha$^{*}$}
\author{Wolfgang Rugel$^{*}$}
\author{Enzo Orlandini$^{*,\dag}$}
\author{Marco Baiesi$^{*,\dag}$}

\address{$*$ Dipartimento di Fisica e Astronomia ``Galileo Galilei'',
 Universit\`a di Padova, Via Marzolo 8, 35131, Padova, Italy
} 
\address{$\dag$ INFN, Sezione di Padova, Via Marzolo 8, 35131, Padova,
Italy}

\date{\today}

\begin{abstract}
By characterising the time evolution of COVID-19 in term of its ``velocity'' (log of the new cases per day) and its rate of variation, or ``acceleration'', we show that in many countries there has been a deceleration even before lockdowns were issued. This feature, possibly due to the increase of social awareness, can be rationalised by a susceptible-hidden-infected-recovered (SHIR) model introduced by Barnes, in which a hidden (isolated from the virus) compartment $H$ is gradually populated by susceptible people, thus reducing the effectiveness of the virus spreading. 
By introducing a partial hiding mechanism, for instance due to the 
impossibility for a fraction of the population to enter the hidden state, we obtain a  model 
that, although still sufficiently simple, faithfully reproduces the different deceleration trends observed in several major countries.
\end{abstract}

\noindent{\it Keywords\/}: epidemic modelling, differential equations, COVID-19

\maketitle

The spread of COVID-19 in all countries is being reported with a massive wealth of data. Although
data are collected by different sources, at different stages and with heterogeneous protocols 
(different national policies, etc.), this huge amount of information allows detailed statistical analysis of the process (see for example~\cite{li20,gatt20,lave20,fane20,carl20,dell20,gaeta20,prib20,gril20,pluc20,carc20,barn20,chen20a,chen20b,thur20,maier20,cast20,prem20,flax20}) and enables robust tests on the universal behaviour of the epidemic dynamics through different communities.

Simple models as the susceptible-infectious-recovered (SIR) scheme~\cite{kerm1927} historically have been used to describe the salient properties of the spreading statistics by relying on an economic amount of parameters and a mean field approach based on a system of nonlinear ordinary differential equations. The SIR model is especially suited to describe closed, spatially homogeneous communities~\cite{murray}. Deviations from the simple SIR dynamics may reveal changes in the social behaviour that eventually reduce the spreading of the disease.

%%%%%%%%%%%%%%%%%%%
\begin{figure}[!b]
 \includegraphics[width=0.56\textwidth]{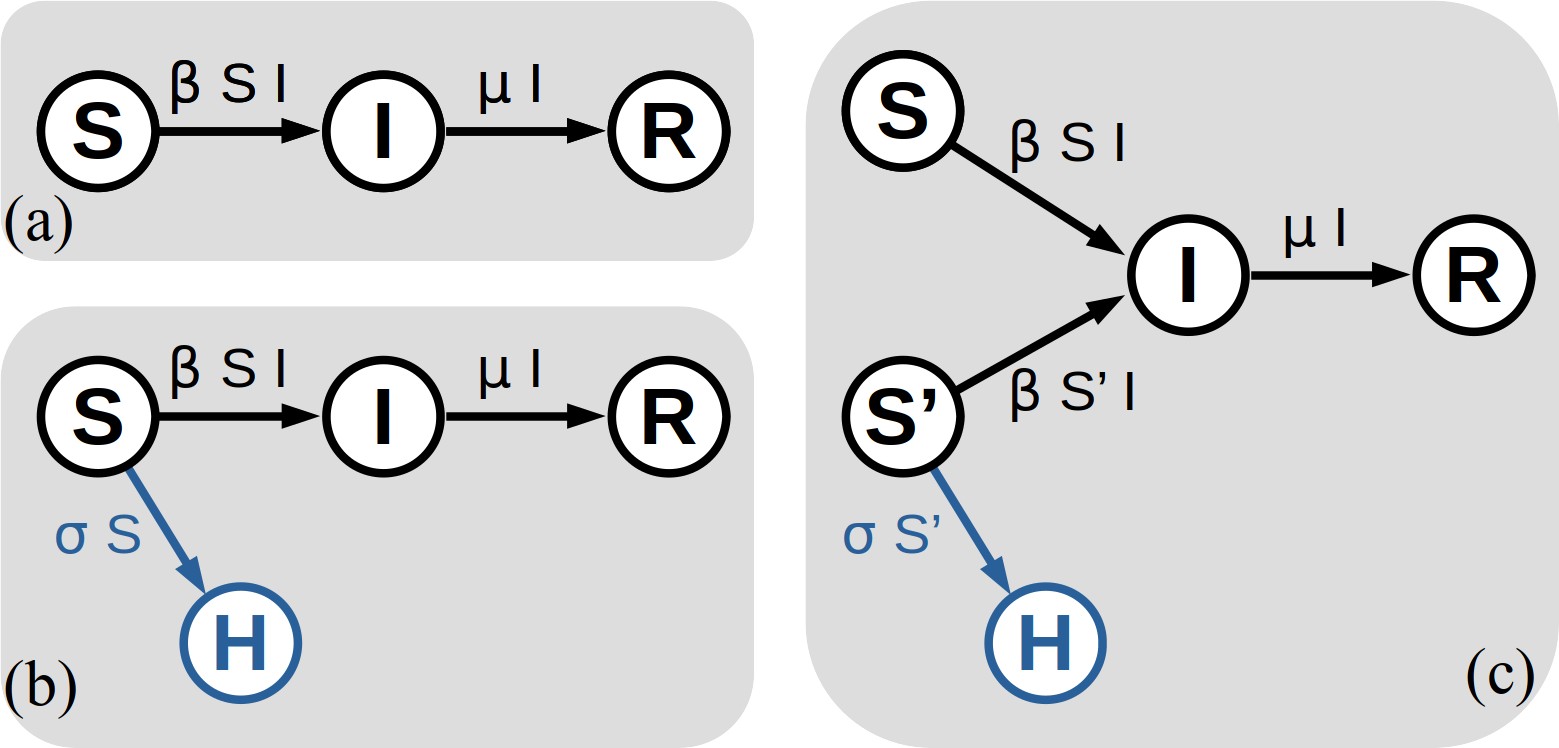}
 \caption{Scheme of each of the models considered in this work: (a) classic SIR, (b) SHIR~\cite{barn20}, and (c) our SS'HIR.}
 \label{fig:mod}
\end{figure}
%%%%%%%%%%%%%%%%%%%

The SIR model may be cast in terms of fractions of population at a given time (day) $t$: $S_t$, $I_t$, and $R_t$, represent respectively the fraction of susceptible, infected, and recovered people, so that $S_t + I_t + R_t = 1$ (note that 
in $R_t$ we are not distinguishing the kind of exit from the infectious state). Its characteristic feature is that susceptibles are infected at a rate proportional both to their number and to the fraction of infected people. By indicating time derivatives by dots, e.g.~$\tder S_t = d S_t/dt$, the SIR evolution sketched in figure\ref{fig:mod}(a) is described by three coupled differential equations,
\begin{align}\label{SIR}
  \begin{cases}    
    \tder S_t  = -\beta I_t S_t \\
    \tder I_t  = \beta I_t S_t - \mu I_t \\
    \tder R_t  = \mu I_t
  \end{cases}
\end{align}
where the ``contact rate'' $\beta$ is a constant determining the strength of the spreading in the transition rate from $S$ to $I$. The healing rate $\mu$, according to the literature, should be related to a healing time $1/\mu$ of the order of two weeks (reported recovery times range from a few days~\cite{faes20} to almost four weeks~\cite{barm20}).
If $\beta>\mu$, the SIR model predicts an exponential explosion of $I_t\sim e^{(\beta-\mu) t}$ in the early stages of the epidemic.
To a good degree, this is the kind of scaling that one could expect in all countries before their lockdown was eventually issued. However, we will show that this is not always the case.

In this paper we collect data for some countries where the statistics is significant and sufficiently regular, especially in the first stage of the pandemic expansion, and we introduce a non-standard way of describing its time evolution, by defining a ``velocity'' $v_t$ of the spreading and its time derivative, or ``acceleration'' $a_t$. A $v$-$a$ diagram is useful for comparing the various stages of the epidemic as it shows at a glance not only the number of new cases per day, but also the trend in its variation, in a range of unities for $v$ and $a$.

From the $v$-$a$ diagrams of several countries it emerges that the COVID-19 spreading was decelerating already before the application of the lockdown. Since the SIR model predicts a constant acceleration, it cannot explain this observed scaling. However, this is fairly well reproduced by a simple extension called SHIR (susceptible-hidden-infectious-recovered) model, recently introduced by Barnes~\cite{barn20}.
In this model one includes a fraction of ``hidden'' people $H_t$, who either decide or are forced to isolate from the rest of the 
susceptible community $S$ at hiding rate $\sigma$, see figure~\ref{fig:mod}(b). The time evolution of SHIR is then governed by the following set of differential equations:
\begin{align}
  \begin{cases}    
 \tder S_t  = -\beta I_t S_t -\sigma S_t \\
 \tder H_t  = \sigma S_t \\
 \tder I_t  = \beta I_t S_t - \mu I_t \\
 \tder R_t  = \mu I_t
 \label{SHIR}
 \end{cases}    
\end{align}
In this picture the deceleration of the epidemic spreading before a lockdown is simply due to a steadily increase 
of the subset $H$ the population that, although susceptible, is becoming aware of the potential danger of the virus and 
changes its social behaviour accordingly.
\rev{The importance of social awareness on epidemic spreading can be investigated by different perspectives. For instance one can look at how the interplay between individual risk perception and resource support between individuals can impact on the epidemic dynamics~\cite{chen20a}.  
Moreover, since individuals react differently to risk,  the effect of heterogeneity in 
self-awareness can be taken into account by considering epidemic spreading dynamics  
on artificial networks and look for the correlation between node degree and self-awareness~\cite{chen20b}.
Our study thus adds to the list of recent works~\cite{barn20,maier20,chen20a,chen20b,thur20,cast20} highlighting that the standard SIR exponential scaling law is not suitable to describe the COVID-19 epidemic.
}

We show below that the SHIR model has indeed an acceleration $a_t$ with an exponential decrease down to 
an asymptotic value. However, although data (before and) after lockdowns comply fairly well with its prediction,
we find that a simple extension of the SHIR model, where only part of susceptible population can access the hidden state,
can improve significantly the fit of the available data. 
This holds both for countries where the lockdown reduced the epidemic spreading to a negative acceleration (e.g.~Germany and Italy) and for those where the lockdown only reduced the acceleration to a positive value smaller than the initial one (e.g.~Brazil and India, during the initial months of the epidemic).
\rev{The fact that a complete depletion of the susceptible compartment is not fully compatible 
with epidemic data was previously discussed in~\cite{cast20} where this possibility was avoided 
by including a reversible hiding mechanism with an additional flux $H\to S$ rather then splitting the susceptible population in $S$ and $S'$ as done here.}

\rev{By considering SIR-like models we thus assume that the population is well mixed. In this scenario 
mean field models are known to give a good description of the epidemic. Note, however, that non-exponential epidemic 
trends might be generated by considering the epidemic dynamics to take place on a quenched 
network describing personal contacts~\cite{thur20}.
}

\section{Data analysis, averaging and rescaling}

Since universal features of disease spreading should better emerge from relative figures, we perform the analysis on the number of confirmed cases over the total population of a given country.
Data were downloaded from the repository for the 2019 Novel Coronavirus Visual Dashboard operated by the Johns Hopkins University Center for Systems Science and Engineering (JHU CSSE)~\cite{data_covid}.

The dataset includes the cumulative number $F_t$ of cases at date $t=0,1,2,\ldots$, i.e.~the total number of people tested to have been infected by the virus up to time $t$, out of $N$ people in the ensemble. In the notation of the SIR model, this number $F_t = N(I_t + R_t)$ corresponds to the current count of infectious people $N I_t$ (databases might add tested asymptomatic people in this counting) plus the 
number $N R_t$ of previously infected people.

%%%%%%%%%%%%%%%%%%%
\begin{figure}[!tb]
 \includegraphics[width=0.78\textwidth]{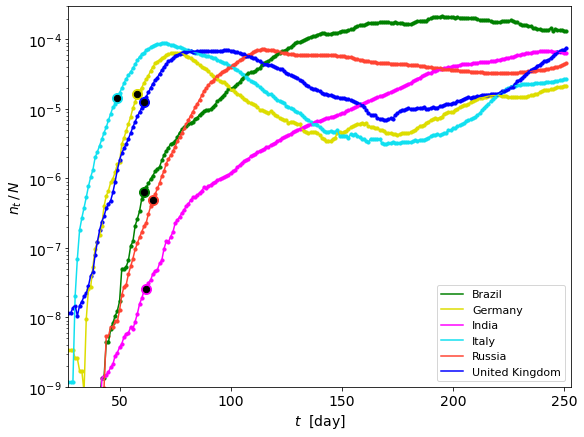}
 \caption{Time series of the number of new cases per day (with $L=14$ days), over the population $N$, for six countries with major COVID-19 outbreaks. Day zero corresponds to Jan. 22nd, 2020, and national lockdowns start at the marked black dots~\cite{lockdown_covid}.}
 \label{fig:t_n}
\end{figure}
%%%%%%%%%%%%%%%%%%%

The new cases per day, $n_t = F_t-F_{t-1} \simeq \tder F_t$ are a manifestation of the epidemic spreading speed.
Since this quantity is noisy due to statistical fluctuations and other factors (variable medical protocols, number of tests, weekly periodicity, etc.) we consider instead the number of new cases occurred in a period ending at time $t$ and averaged (smoothed) over a window of $L$ days:
\begin{align}
 n_t^{(L)} = \frac{ F_t - F_{t-L}}L.
 \label{eq1:n}
\end{align}
This quantity is reported in figure~\ref{fig:t_n} for different countries.
The figure shows that, over a time scale of two to three weeks, after lockdowns (black dots) there has been either a decrease of the number of cases per day or a decrease in the acceleration of the spreading (smaller positive slope of curves). In later periods, when the lockdowns were removed in the European counties, one notes the restart of second and further waves of the epidemic in those countries. However, maybe less evident, there also appears to be some negative curvature in the plots before lockdowns. This feature is better analysed with the following procedure.

A comparison to a time-shifted $n_{t-z}$ gives the variation
\begin{align}
 \Delta n_t^{(L,z)} = {n_t^{(L)} - n_{t-z}^{(L)}},
 \label{eq2:n}
\end{align}
where $z=1$ is the minimum value.
In presence of an exponential growth, we thus get a smoothed estimate of the daily rate of increase $\Delta n_t^{(L,1)} \simeq e^{\beta-\mu}$, which is the exponential of the infection rate of the SIR model.
In general we should have $\Delta n_t^{(L,z)}\simeq [\Delta n_t^{(L,1)}]^z$.
From now on we consider $L=14$, $z=3$.

\begin{figure}[!tb]
 \includegraphics[width=\textwidth]{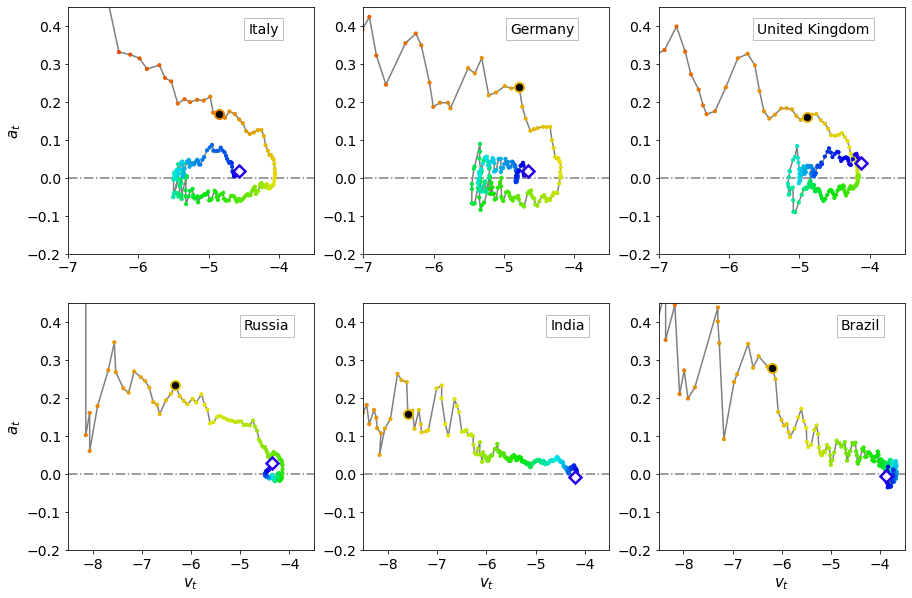}
 \caption{Acceleration vs velocity ``falling star'' trajectories for six large countries with major COVID-19 outbreaks: time parameterises trajectory with a colour code from red to blue, black dots mark the major lockdown dates during Spring 2020, and an empty diamond indicates the last day of a trajectory.
  Starting from the upper left corner, trajectories continue to increase the spreading velocity $v_t$ as long as $a_t$ remains positive while decreasing. When $a_t$ turns to negative values, trajectories start to regress the velocity to lower values. It is visible that when social measures are loosened (green turning to blue region for Italy, Germany, and UK), fluctuations seem to bring back the trajectories to the $a_t>0$ phase, leading to oscillations around the $a\approx 0$ region or to eventual further outbreaks.}
 \label{fig:v_a}
\end{figure}

The typical exponential trends of the disease evolution suggest to consider log scales for a better visualisation and characterisation of the stage of the spreading.
Hence, we portray the trajectory of the disease (parameterised by time $t$) by its velocity $v_t$ in log scale 
and by the rate of its variation, or ``acceleration'' $a_t$. These are defined by their proxies
\begin{align}
 v_t & = \log_{10}[ n_t^{(L)} / N]\\
 a_t & = \frac 1 z \frac{\Delta n_t^{(L,z)}}{(n_t^{(L)}+n_{t-z}^{(L)})/2}
 =
 \frac 1 z \frac{n_t^{(L)} - n_{t-z}^{(L)}}{(n_t^{(L)}+n_{t-z}^{(L)})/2}
 \simeq
 (\ln 10)\tder v_t
 \label{eq3:n}
\end{align}
where $N$ is the total population of a given country. The numerical pre-factor $\ln 10$ in the definition of the acceleration  simplifies formulas later.
Figure~\ref{fig:v_a} shows the trajectories of the epidemic spreading in the $(v,a)$ plane for the six selected countries: one can notice that all trajectories start from the upper left corner, they continue with increasing $v$ ($a>0$) and eventually cross 
the $a=0$ axis. From there, typically $a$ becomes negative indicating a period of recession of the viral disease. 
However, this occurs with such a slowly receding velocity (small negative values of $a$) that random fluctuations in the population of $S$ or $I$ may bring the system back to the dynamics characterised by $a>0$, i.e $v$ increasing phase. Note that 
for some countries as Brazil and India, the period of positive acceleration has been longer than in other countries and recession has been, so far, negligible. It is particularly for these cases that an extension of the SHIR model is needed (see next section). 

%%%%%%%%%%%%%%%%%%%
%%%%%%%%%%%%%%%%%%%

Clearly all trajectories show a consistent drop in $a_t$ after the application of the national lockdowns (see black point along the trajectory). Nevertheless, there are countries in which deceleration sets in several days before their lockdowns due to the application of some preventive not draconian measures. In Germany, for instance, the cancellation of several public events started before the official lockdown and has probably slowed down the spreading of COVID-19. In Italy, the national lockdown was preceded by the isolation of so called local "red zones" (where initial cases were detected) from the rest of the community.
This measure would be placed at the beginning of the trajectory in figure~\ref{fig:v_a}, suggesting that it was effective in inducing the strong initial decrease of the spreading acceleration in Italy.

Next, we analyse this non trivial behaviour analytically for the SIR and SHIR model, then we perform fits to data to assess the performance of the SHIR model, finally we extend this model to improve the agreement with data 
 and we discuss the implications of our results.

\section{Model trends of the acceleration}

Since the simplest part of the epidemic evolution ranges from its initial rise to the end of the lockdowns, we have isolated 
the portion of the time series within this time window. For each country this is done by shifting the time $t-t_0 \to t$ to set $t=0$ to a day ($t_0$ in the original scale) close to the first maximum of $a_t$ and by keeping a period of about two to three months since then. In this way we 
exclude the possibility of including in the analysis the onset of potential second waves of the spreading.

Let us start by considering the SHIR dynamics \eqref{SHIR}, of which the standard SIR model is a special case with $\sigma=0$.
For this model the confirmed cases have a time derivative
\begin{align}
\tder F_t = N(\tder I_t + \tder R_t) = N \beta S_t I_t
\end{align}
and the velocity becomes
\begin{align}
 v & = \log_{10} \tder F_t / N\non
 & = \log_{10} \beta + \log_{10} S_t + \log_{10} I_t.
\end{align}
Since the fraction of infected people has been always at most of the order of $10^{-4}$ one can safely assume that
the decreasing of the susceptible population $S_t$ is mainly due
to an increase of the hidden fraction of responsible people, $H_t$. If the rate $\sigma$ is large enough, i.e.~so that $\sigma S_t \gg \beta S_t I_t$, from (\ref{SHIR}) it follows that $S_t$ decays exponentially as $S_t = S_0 e^{-\sigma t}$. Using this scaling, we have a velocity
\begin{align}
 v_t & \simeq \log_{10} \beta + \log_{10} S_0 - \frac{\sigma t}{\ln  10} + \log_{10} I_t,
 \label{SHIR_v}
\end{align}
and acceleration 
\begin{align}
 a_t \simeq \beta S_0 e^{-\sigma t} - \mu - \sigma.
 \label{SHIR_a}
\end{align}

%%%%%%%%%%%%%%%%%%%
\begin{figure}[!tb]
 \includegraphics[width=\textwidth]{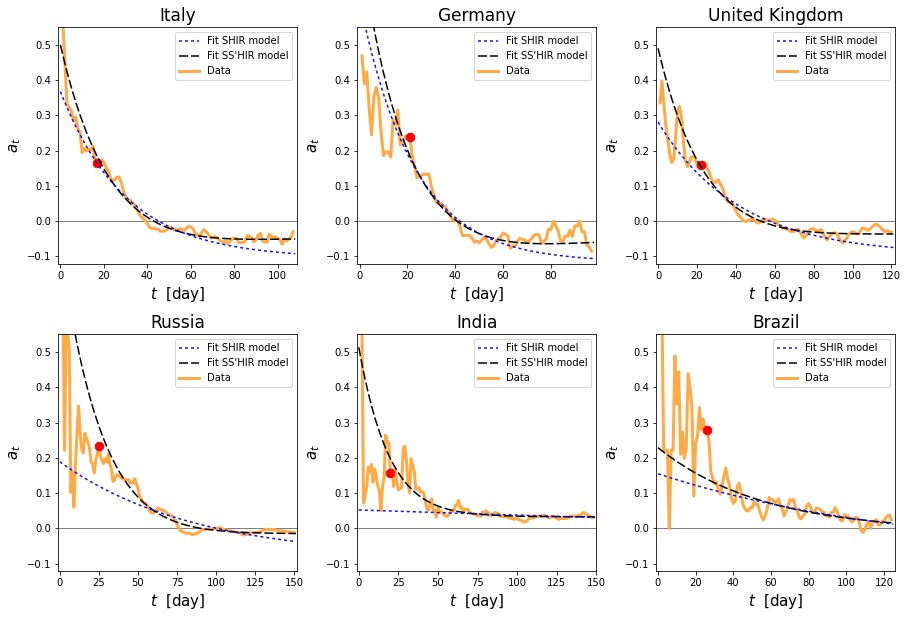}
 \caption{Time series of the virus spreading ``acceleration'' $a_t$ for six countries with major COVID-19 outbreaks, in the period before eventual second waves and with red dots marking the lockdown dates, and fits according to SHIR and SS'HIR models, both performed with dedicated python libraries for nonlinear fits. For Italy, Germany, UK, and Russia one notes that the asymptotic value of $a_t$ is not correctly recovered by the SHIR model. This value is only approximated for India and Brazil by introducing a tiny value $\sigma$ for the transition rate to the hidden state (see table~\ref{tab:1}).
The SS'HIR variant, on the other hand, can fit the long time value as well as the early stage of $a_t$ by yielding reasonable values of $\sigma$. 
 }
 \label{fig:t_a}
\end{figure}
%%%%%%%%%%%%%%%%%%%

Hence, according to the SIR model, since in most countries the condition $F_t\ll N$ is still fulfilled and $S_t\simeq 1$, one would expect a spreading with constant acceleration $a\simeq \beta - \mu$ from \eqref{SHIR_a}.
This contrasts with data shown in figure~\ref{fig:v_a} and figure~\ref{fig:t_a}: even at initial stages (before lockdowns), in Germany, Italy, and UK it is readily seen a {\em natural} decrease of $a_t$. In this respect the SIR model is too simple to 
describe the initial trend of the COVID-19 spreading, as highlighted also by Barnes~\cite{barn20} by simply looking at the ratio of new cases over the total ones. 

On the other hand the SHIR model explains economically the deceleration by allowing a fast discharge of $S_t$ via a sufficiently large hiding rate $\sigma$; this is a sensible assumption, as it does not take many days to organise a reduced level of contact between people.

%%%%%%%%%%%%%%%%%%%%%%%%%%%%%%%%%%%%%%
\begin{table}[b!]
\caption{\label{tab:1}Initial time $t_0$ (days from Jan. 22nd, 2020) and parameters from fits of the acceleration $a_t$ with the SHIR model and with the SS'HIR model, shown in figure~\ref{fig:t_a}. Parameters $\beta$ and $\sigma$ are in day$^{-1}$ units.}
\begin{indented}
\item[]\begin{tabular}{lr|rr|rrr|}
 \toprule
 & \multicolumn{1}{c}{}&\multicolumn{2}{c}{SHIR}&\multicolumn{3}{c}{SS'HIR}\\
   \cline{3-7}
{} & $t_0$ & $\beta$ & $\sigma$ & $\beta$ & $\sigma$ & $p$\\
\midrule
Italy          & 15 & 0.47 & 0.033 & 0.61 & 0.046 & 0.961 \\
Germany        & 20 & 0.74 & 0.044 & 0.92 & 0.054 & 0.981 \\
United Kingdom & 22 & 0.38 & 0.024 & 0.60 & 0.042 & 0.941 \\
Russia         & 23 & 0.27 & 0.012 & 0.93 & 0.038 & 0.939 \\
India          & 25 & 0.13 & 0.001 & 0.63 & 0.049 & 0.833 \\
Brazil         & 18 & 0.23 & 0.008 & 0.31 & 0.017 & 0.798 \\
\bottomrule
\end{tabular}
\end{indented}
\end{table}
%%%%%%%%%%%%%%%%%%%%%%%%%%%%%%%%%%%%%%

In figure~\ref{fig:t_a} we show the time series of the epidemic acceleration  for different countries. The dotted curves
represent the fits of data based on \eqref{SHIR_a} where we have assumed $S_0\lesssim 1$,  $\mu=1/14$ day$^{-1}$ and let 
the parameters $\sigma$ and $\beta$ vary freely (the tiny initial $I_0$ value is also left as a free parameter). Note that it is not possible to consider $\mu$ as a further free parameter since the number of new cases per day is not very sensible to its value. Fortunately, this also means that the fits are quite independent on the chosen value of $\mu$. By assigning a weight $w_t = n_t$ to each data point, the best fit gives the estimates of $\sigma$ and $\beta$ reported in table~\ref{tab:1}.
In figure~\ref{fig:t_a} one notes that the SHIR model (see dotted curves) gets sufficiently close to the observed 
trends of the acceleration; there are however some visible deviations at short times and, most importantly, in some cases also at long times. Furthermore, some very small estimates of $\sigma$ reported in table~\ref{tab:1} corroborate the 
idea that some extension of the SHIR model may improve the agreement with the real data.

For these reasons, we test a model in which the long time limit of $S_t$ is not negligible even if $\sigma\ne 0$. This is obtained by dividing the fraction of susceptible population in two groups: a fraction $S'$ can access freely the hidden state, as sketched in figure~\ref{fig:mod}(c), while a fraction $S$ could not access the isolated condition (for reasons as work duties, poverty, etc.).
The evolution equations of this SS'HIR model become
\begin{align}
  \begin{cases}
  \tder S_t  = -\beta I_t S_t \\
  \tder S'_t  = -\beta I_t S'_t -\sigma S'_t \\
  \tder H_t  = \sigma S'_t \\
  \tder I_t  =  \beta I_t (S_t+S'_t) - \mu I_t \\
  \tder R_t  =  \mu I_t
  \label{SSHIR}
  \end{cases}
\end{align}
To partition the initial fraction of the susceptibles we consider a  parameter $p\le 1$ such that $S'_0=p (1-I_0)\simeq p$ and $S_0 =(1-p) (1-I_0)\simeq 1-p$.
The number of new cases per day in this model is given by
\begin{align}
\dot{F}_t = N(\dot{I}_t+\dot{R}_t) = N \beta (S_t + S'_t) I_t
\end{align}
and the velocity becomes
\begin{align}
  v_t = \log_{10} \dot{F}/N = \log_{10}\beta + \log_{10}(S_t + S'_t) + \log_{10}I_t.
\end{align}
By assuming $S_t \simeq S_0$ and $S'_t \simeq S'_0e^{-\sigma t}$ we thus get
\begin{align}
  v_t &\simeq \log_{10} \dot{F}/N = \log_{10}\beta + \log_{10}(S_0 + S'_0 e^{-\sigma t}) + \log_{10}I_t\\
  a_t &\simeq \frac{(-\sigma)S'_0 e^{-\sigma t}}{S_0 + S'_0 e^{-\sigma t}} + \frac{\dot{I}_t}{I_t}.
\end{align}
Since $\dot{I}_t/I_t = \beta (S_t + S'_t) - \mu$, the acceleration reduces to
\begin{align}
  a_t &\simeq \frac{(-\sigma)S'_0 e^{-\sigma t}}{S_0 + S'_0 e^{-\sigma t}} + \beta (S_0 + S'_0e^{-\sigma t}) - \mu \non
     &\simeq \frac{(-\sigma)p e^{-\sigma t}}{1-p + p e^{-\sigma t}} + \beta (1-p + p e^{-\sigma t}) - \mu.
\end{align}
By fitting data with this formula we get the values listed in table~\ref{tab:1}, which are
more realistic than those obtained  from the  fits based on  the SHIR model. For instance, the unrealistic value 
$\sigma \simeq 0.001$ day$^{-1}$ now becomes $\sigma \simeq 0.049$ day$^{-1}$, in line with other values.
The fits in figure~\ref{fig:t_a} show that the SS'HIR model can correctly capture the trend of the acceleration in a long time span, including also long time values of $a_t\approx 0$ (Russia) or $a_t>0$ (Brazil and India).~\footnote{By ``long time'' we mean an intermediate regime in which the lockdown has generated stable conditions.}
Note that the estimated values of  $p\simeq 0.8$ for Brazil and India mean that, in the SS'HIR modelling, a fraction $S_0 \simeq 1-p\approx 0.2$ of the population cannot isolate itself from the rest of the community. A classic SIR dynamics on this $20\%$ of the population, with  $\beta (1-p) - \mu \gtrsim 0$ might be the reason for which the epidemic has been continuing to accelerate for some months in those countries.

\begin{figure}[!tb]
 \includegraphics[width=\textwidth]{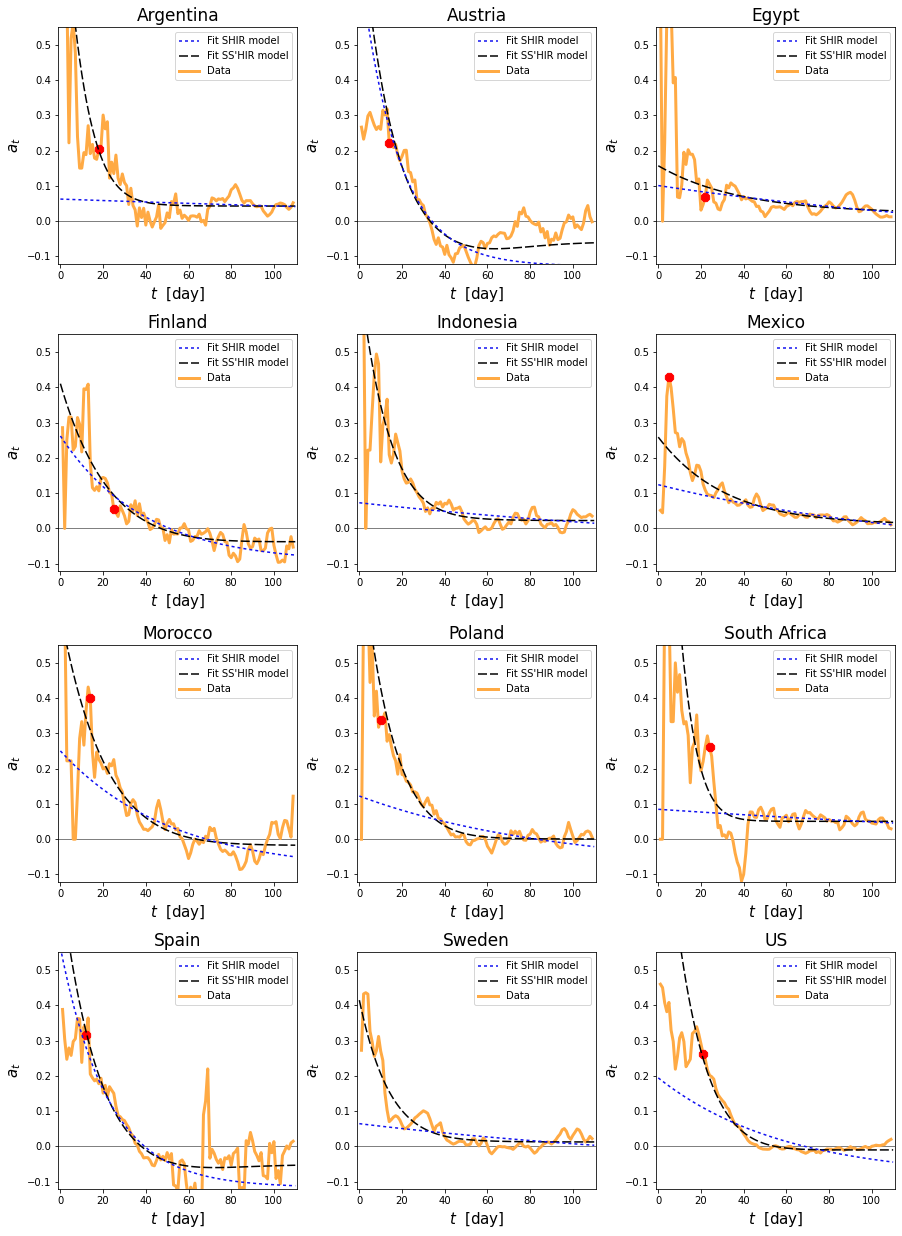}
 \caption{As in figure~\ref{fig:t_a}, for twelve different countries.
 }
 \label{fig:t_a2}
\end{figure}

\rev{The analysis of twelve more countries 
(see figure~\ref{fig:t_a2} and table~\ref{tab:2}) shows that
the SS'HIR model is flexible enough to fit different scenari. For example 
it reproduces fairly well both the non-monotonic trend of $a_t$ for Austria and  the convergence to 
the almost null value of  $a_t$ in United States, Sweden, and Poland 
(whose possible origin is discussed in~\cite{thur20}).}

%%%%%%%%%%%%%%%%%%%
%%%%%%%%%%%%%%%%%%%

%%%%%%%%%%%%%%%%%%%%%%%%%%%%%%%%%%%%%%
\begin{table}[t!]
\caption{\label{tab:2} As in table~\ref{tab:1}, for twelve different countries. The initial time is $t_0=23$ days, besides $t_0=30$ days for Mexico.}
\begin{indented}
\item[]\begin{tabular}{l|rr|rrr|}
 \toprule
 &\multicolumn{2}{c}{SHIR}&\multicolumn{3}{c}{SS'HIR}\\
   \cline{2-6}
{}  & $\beta$ & $\sigma$ & $\beta$ & $\sigma$ & $p$\\
\midrule
Argentina &  0.14 &  0.002 &  1.22 &  0.091 &  0.906 \\
Austria   &  0.86 &  0.056 &  1.02 &  0.065 &  0.987 \\
Egypt     &  0.18 &  0.005 &  0.24 &  0.024 &  0.619 \\
Finland   &  0.36 &  0.026 &  0.52 &  0.043 &  0.934 \\
Indonesia &  0.15 &  0.004 &  0.85 &  0.069 &  0.890 \\
Mexico    &  0.20 &  0.007 &  0.35 &  0.030 &  0.769 \\
Morocco      &  0.34 &  0.019 &  0.76 &  0.046 &  0.929 \\
Poland       &  0.20 &  0.011 &  1.00 &  0.065 &  0.928 \\
South Africa &  0.16 &  0.003 &  2.18 &  0.112 &  0.944 \\
Spain        &  0.68 &  0.045 &  0.88 &  0.057 &  0.976 \\
Sweden       &  0.14 &  0.005 &  0.54 &  0.064 &  0.844 \\
US           &  0.28 &  0.017 &  1.33 &  0.067 &  0.954 \\
\bottomrule
\end{tabular}
\end{indented}
\end{table}
%%%%%%%%%%%%%%%%%%%%%%%%%%%%%%%%%%%%%%

\rev{Finally, we notice that the estimated values of the contact rate $\beta$ are likely to be  an overestimate of the 
parameter $\beta_{\rm true}$ that corresponds to an  expanded scenario in which asymptomatic (i.e.~undetected infected) cases are taken into account~\cite{prib20,gaeta20,li20,gatt20,lave20}.
Indeed, in the simplest possible description, the main effect of these asymptomatic cases on the early 
epidemic dynamics might be a biased estimate $\beta = \beta_{\rm true} / f$, where $f$ is the probability of a susceptible person to become infected rather than asymptomatic. This can be shown 
  by mapping a SIR dynamics with $\beta_{\rm true}$ and healing rate $\mu$, in which visible infected cases $I_t$ and asymptomatic ones $A_t$ are summed within a compartment, to the SIR dynamics without $A_t = \frac{1-f}f I_t$.
These two SIR models yield the same depletion of $S_t$ if their constant are related as $\beta = \beta_{\rm true} / f$.
Note that $\beta_{\rm true} < \beta$ and its value may depend (through $f$) also on the efficiency of a given health care system to detect infected cases.
Assuming a fraction $f\approx 0.5$~\cite{lave20}, we would have $\beta_{\rm true}$ around one half of the values shown in table~\ref{tab:1}.
  Possibly the presence of asymptomatic cases will emerge more clearly in the late stage of the epidemic dynamics, when $S_t=1-I_t-A_t-R_t$ becomes sufficiently different from the measured non infected fraction $1-I_t-R_t$, so that herd immunity is reached earlier than expected.
 }

\section{Discussion}

We have presented a velocity-acceleration diagram that visualises the epidemic state and its trend.
In the $v$-$a$ diagrams of COVID-19 evolution, illustrated in this work for six large countries, we note that the acceleration is not constant often in periods including days before national lockdowns. The observed deceleration cannot be explained by a bias introduced by a variable number of tests, which were in general following an opposite trend increasing with time.
This suggests that social distancing, either introduced by local lockdowns, personal choices, or cancellation of public events due to the news from Asia, was already effectively reducing to some degree the spreading of the virus.

The simplest effective explanation of the observed deceleration in the number of new COVID-19 cases per day comes from assuming that the fraction of susceptible population is reduced over a time scale $1/\sigma$ of tens of days/weeks by the isolation imposed by the national lockdowns. This confirms the findings by Barnes' with the SHIR model. Building upon this model, we have introduced a simple modification in which only part of the population can comply with the enforcement of strict social distancing. The remaining part obeys the usual rules of the SIR model. Our modification better fits the data from several countries, especially those where the acceleration of the spreading has remained positive for months, where our fits suggest that about $20\%$ of the population effectively was never socially isolated.

There remains to understand how much the timescale $1/\sigma$ that emerges for most of the isolation dynamics is actually representative of a more complex mixture of effects, eventually including the personal evolution in the stages of the illness (as for exposed compartments in the SEIR model~\cite{carc20}), which is not considered in the basic SIR model.
However, the SEIR and similar variants of the SIR model could not fit quantitatively, so far, 
the observed deceleration of the pandemic spreading (data not shown) if a fraction of “hidden” people $H$ is not included.
\rev{Moreover, also simple models including an asymptomatic fraction of the population could not explain 
the observed deceleration of the epidemic: we have argued that the effect of the undetected cases in the early epidemic is mostly just a rescaling of the measured contact rate $\beta$.}

\section*{Acknowledgements}
We acknowledge useful discussions with Francesco Piazza and Samir Suweis.
This work was initiated and carried out in large part by the first three authors as a project in the {\em Physics of Data} Master's Degree course at the University of Padova.

\section*{References}

%%\bibliography{bib_virus}

\begin{thebibliography}{10}
\expandafter\ifx\csname url\endcsname\relax
  \def\url#1{{\tt #1}}\fi
\expandafter\ifx\csname urlprefix\endcsname\relax\def\urlprefix{URL }\fi
\providecommand{\eprint}[2][]{\url{#2}}

\bibitem{li20}
Li R, Pei S, Chen B, Song Y, Zhang T, Yang W and Shaman J 2020 {\em Science\/}
  {\bf 368} 489--493 (\textit{Preprint}
  \eprint{https://science.sciencemag.org/content/368/6490/489.full.pdf})
  \urlprefix\url{https://science.sciencemag.org/content/368/6490/489}

\bibitem{gatt20}
Gatto M, Bertuzzo E, Mari L, Miccoli S, Carraro L, Casagrandi R and Rinaldo A
  2020 {\em Proc. Natl. Acad. Sci.\/} {\bf 117} 10484--10491 (\textit{Preprint}
  \eprint{https://www.pnas.org/content/117/19/10484.full.pdf})
  \urlprefix\url{https://www.pnas.org/content/117/19/10484}

\bibitem{lave20}
Lavezzo E, {\em et al}, Crisanti A and {Imperial College {COVID}-19 Response
  Team} 2020 {\em Nature\/} {\bf 584} 425--429
  \urlprefix\url{https://www.nature.com/articles/s41586-020-2488-1}

\bibitem{fane20}
Fanelli D and Piazza F {2020} {\em Chaos Solit. Fract.\/} {\bf {134}} {109761}

\bibitem{carl20}
Carletti T, Fanelli D and Piazza F 2020 {\em Chaos Solit. Fract.: X\/} {\bf 5}
  \urlprefix\url{https://www.sciencedirect.com/science/article/pii/S2590054420300154}

\bibitem{dell20}
Dell’Anna L 2020 {\em Scientific Reports\/} {\bf 10} 15763
  \urlprefix\url{http://dx.doi.org/10.1038/s41598-020-72529-y}

\bibitem{gaeta20}
Gaeta G 2021 {\em Mathematics in Engineering\/} {\bf 3} 1–39
  \urlprefix\url{http://dx.doi.org/10.3934/mine.2021013}

\bibitem{prib20}
Pribylova L and Hajnova V 2020 {SEIAR} model with asymptomatic cohort and
  consequences to efficiency of quarantine government measures in {COVID}-19
  epidemic (\textit{Preprint} \eprint{arXiv:2004.02601})

\bibitem{gril20}
Grilli J, Marsili M and Sanguinetti G 2020 Estimating the impact of preventive
  quarantine with reverse epidemiology (\textit{Preprint}
  \eprint{arXiv:2004.04153})

\bibitem{pluc20}
Pluchino A, Biondo A~E, Giuffrida N, Inturri G, Latora V, Moli R~L, Rapisarda
  A, Russo G and Zappala' C 2020 A novel methodology for epidemic risk
  assessment: the case of {COVID}-19 outbreak in {I}taly (\textit{Preprint}
  \eprint{arXiv:2004.02739})

\bibitem{carc20}
Carcione J~M, Santos J~E, Bagaini C and Ba J 2020 {\em Frontiers in Public
  Health\/} {\bf 8} 230
  \urlprefix\url{https://www.frontiersin.org/article/10.3389/fpubh.2020.00230}

\bibitem{barn20}
Barnes T 2020 The {SHIR} model: Realistic fits to {COVID}-19 case numbers
  (\textit{Preprint} \eprint{arXiv:2007.14804})

\bibitem{chen20a}
Chen X, Liu Q, Wang R, Li Q and Wang W 2020 {\em Complexity\/} {\bf 2020}
  3256415

\bibitem{chen20b}
Chen X, Gong K, Wang R, Cai S and Wang W 2020 {\em Applied Mathematics and
  Computation\/} {\bf 385} 125428 ISSN 0096-3003
  \urlprefix\url{http://www.sciencedirect.com/science/article/pii/S0096300320303891}

\bibitem{thur20}
Thurner S, Klimek P and Hanel R 2020 {\em Proc. Nat. Acad. Sci.\/} {\bf 117}
  22684--22689 ISSN 0027-8424 (\textit{Preprint}
  \eprint{https://www.pnas.org/content/117/37/22684.full.pdf})
  \urlprefix\url{https://www.pnas.org/content/117/37/22684}

\bibitem{maier20}
Maier B~F and Brockmann D 2020 {\em Science\/} {\bf 368} 742--746 ISSN
  0036-8075 (\textit{Preprint}
  \eprint{https://science.sciencemag.org/content/368/6492/742.full.pdf})
  \urlprefix\url{https://science.sciencemag.org/content/368/6492/742}

\bibitem{cast20}
Castro M, Ares S, Cuesta J~A and Manrubia S 2020 {\em Proc. Nat. Acad. Sci.\/}
  {\bf 117} 26190--26196 ISSN 0027-8424 (\textit{Preprint}
  \eprint{https://www.pnas.org/content/117/42/26190.full.pdf})
  \urlprefix\url{https://www.pnas.org/content/117/42/26190}

\bibitem{prem20}
Prem K, Liu Y, Russell T~W, Kucharski A~J, Eggo R~M, Davies N, Jit M, Klepac P
  and COVI C~M~M~I~D {2020} {\em {Lancet Public Health}\/} {\bf {5}}
  {E261--E270}

\bibitem{flax20}
Flaxman S, Mishra S, Gandy A, Unwin H~J~T, Mellan T~A, Coupland H, Whittaker C,
  Zhu H, Berah T, Eaton J~W, Monod M, Ghani A~C, Donnelly C~A, Riley S, Vollmer
  M~A~C, Ferguson N~M, Okell L~C and Bhatt S {2020} {\em {Nature}\/} {\bf
  {584}} {257+}

\bibitem{kerm1927}
Kermack W~O and McKendrick A~G 1927 {\em Proc. R. Soc. Lond.\/} {\bf A115}
  700--721

\bibitem{murray}
Murray J 2002 {\em Mathematical Biology: I. An Introduction\/} third edition ed
  (Springer)

\bibitem{faes20}
  Faes C, Abrams S, {Van Beckhoven} D, Meyfroidt G, Vlieghe E and Hens N 2020
{\em   Int. J. Environ. Res. Public Health} 
{\bf   17}
7560

\bibitem{barm20}
Barman M~P, Rahman T, Bora K and Borgohain C 2020 {\em Diabetes \& Metabolic
  Syndrome: Clinical Research \& Reviews\/} {\bf 14} 1205 -- 1211
  \urlprefix\url{http://www.sciencedirect.com/science/article/pii/S1871402120302502}

\bibitem{data_covid}
Data repository for the 2019 Novel Coronavirus Visual Dashboard operated by the
  Johns Hopkins University Center for Systems Science and Engineering (JHU
  CSSE) \urlprefix\url{https://github.com/CSSEGISandData/COVID-19}

\bibitem{lockdown_covid}
{COVID}-19 Lockdown dates by country
  \urlprefix\url{https://www.kaggle.com/jcyzag/covid19-lockdown-dates-by-country}

\end{thebibliography}

\providecommand{\newblock}{}

\end{document}